\def\diff#1#2{\frac{\partial #1}{\partial #2}}
\def\tf{t_s(r)}

\documentclass[oneside,a4paper,10pt]{amsart}
\usepackage{amsmath}               
\usepackage{amscd}
\usepackage{amsthm}                
\usepackage{times}

\numberwithin{equation}{section}

\title[New mathematical framework for spherical gravitational
collapse]%
{New mathematical framework for spherical gravitational collapse}

\author[R.\ Giamb\`o ,\ F.\ Giannoni]{Roberto Giamb\`o, Fabio Giannoni}
\address{Dipartimento di Matematica e Informatica,
Universit\`a di Camerino, Italy}
\email{roberto.giambo@unicam.it, fabio.giannoni@unicam.it}
\author[G.\ Magli]{Giulio Magli}
\address{Dipartimento di Matematica, Politecnico di
Milano, Italy}
\email{magli@mate.polimi.it}
\author[P.\ Piccione]{Paolo Piccione}
\address{Dipartimento di Matematica e Informatica,
Universit\`a di Camerino, Italy \hfill\break\indent
(on leave from Departamento de Matem\'atica,\hfill\break\indent
Universidade de S\~ao Paulo, Brazil)}
\email{paolo.piccione@unicam.it, (piccione@ime.usp.br)}

\begin{document}
\swapnumbers
\theoremstyle{plain}\newtheorem{teo}{Theorem}[section]
\theoremstyle{plain}\newtheorem{prop}[teo]{Proposition}
\theoremstyle{plain}\newtheorem{lem}[teo]{Lemma}
\theoremstyle{plain}\newtheorem{cor}[teo]{Corollary}
\theoremstyle{definition}\newtheorem{defin}[teo]{Definition}
\theoremstyle{remark}\newtheorem{rem}[teo]{Remark}
\theoremstyle{plain} \newtheorem{assum}[teo]{Assumption}
\theoremstyle{definition}\newtheorem{example}[teo]{Example}

\begin{abstract}

A theorem, giving necessary and sufficient condition for naked
singularity formation in spherically symmetric non static
spacetimes under hypotheses of physical acceptability, is
formulated and proved. The theorem relates existence of singular
null geodesics to existence of regular curves which are
super-solutions of the radial null geodesic equation, and allows
us to treat all the known examples of naked singularities from a
unified viewpoint. New examples are also found using this
approach, and perspectives are discussed.

\end{abstract}

\maketitle

\begin{section}{Introduction}\label{sec:intro}

The validity of the so called Cosmic Censorship Conjecture, i.e.
the idea that physically acceptable collapsing systems should
always form blackholes is, as is well known, seriously put in
doubt by various counterexamples, that is, exact solutions of the
Einstein field equations exhibiting naked singularities.

Recently, we re-considered this problem in an effort of studying
the validity of censorship, at least in spherical symmetry, with
the application to the geodesic equations of techniques of
non-linear o.d.e. without resorting to exact solutions of
Einstein's. It is, in fact, obvious that exact solutions, despite
their obvious interest, cannot give the complete answer to this
problem due to the non-linearity of the field equations.

Our first results, presented recently \cite{GGM}, adopted the
`censorist' point of view, in that we provided a sufficient
condition for black hole formation in spherical symmetry. However,
the existence of such a condition does not necessarily mean that
the room available for the Censor is very large (that is, that the
hypotheses of the theorem are likely to be realized in generic
situations). In order to face this problem, we take here the
opposite viewpoint and obtain a sufficient condition for naked
singularity formation (the condition, as we shall see, is also
trivially shown to be necessary). The theorem allows us to give a
simple, complete classification of (virtually) all the widespread
zoo of naked singularities existing in the literature, and to
produce new examples as well.

\end{section} 

\begin{section}{The existence of naked singularities}\label{sec:nak}

We consider a spherically symmetric collapsing object (the matter
source can be any model compatible with the weak energy
condition). The general, spherically symmetric, non--static line
element in comoving coordinates $t,r,\theta,\varphi$ can be
written in terms of three functions $\nu , \lambda , R$ of $r$ and
$t$ only as follows:
$$
\mathrm ds^2= -e^{2\nu}\mathrm dt^2 + e^{2\lambda}\mathrm dr^2 + R^2
(\mathrm d\theta^2 +
\sin^2\theta\mathrm d\varphi^2)\ ,
$$

A fundamental quantity is the mass function $m(r,t)$ defined in
such a way that  the equation $R=2m$ spans the boundary of the
{\it trapped region}, i.e. the region in which outgoing null rays
re-converge:
\begin{equation}\label{eq:Psi1}
m(r,t)=(R/2)\left(1- g^{\mu\nu}\partial_\mu R\,\partial_\nu
R\right).
\end{equation}
The curve $t_h(r)$ defined via $R(r,t_h(r))=2m(r,t_h(r))$ is
called {\it apparent horizon}.

We consider here only those matter configurations, which admit a
regular center, and we always suppose that the collapse starts
from regular initial data on a Cauchy surface ($t=0$, say), so
that the singularities forming are a genuine outcome of the
dynamics. If the solution is initially regular, the spacetime can
become singular whenever $R=0$ (focusing singularities) or $R'=0$
(crossing singularities). We consider here only focusing
singularities.

The locus of the zeroes of the function $R(r,t)$ defines the {\it
singularity curve} $\tf $ by the relation $R(r,\tf )=0$.
Physically, $\tf$ is the comoving time at which the shell of
matter labeled by $r$ becomes singular. The singularity forming at
$r=0,t=t_s (0)$ is called {\sl central} as opposed to those
occurring at $r=r_0>0,t=t_s (r_0)$. A singularity cannot be naked
if it occurs after the formation of the apparent horizon. Since
$R$ vanishes at a singularity, any naked singularity in spherical
symmetry must be massless; at a massless singularity the horizon
and the singularity form simultaneously. Since regularity of the
center up to singularity formation requires $m(0,t)=0\,\forall t<
t_s(0)$, the center is always a candidate for nakedness. Usually,
all other points ("non-central points") of the singularity curve
lie after the formation of the apparent horizon, i.e.
$t_h(r)<t_s(r)$ for $r>0$, and the singularity is therefore
covered. This always happens, for instance, if the radial pressure
is positive. Then we concentrate here on the central singularity,
although our results can easily be extended to non-central
singularities as well.

To analyze the causal structure, observe that, if the singularity
is visible to nearby observers, at least one outgoing null
geodesic must exist, that meets the singularity in the past. Such
a geodesic will be a solution of

\begin{equation}\label{f}
\begin{cases}
\frac{\mathrm dt}{\mathrm dr}=\varphi (r,t), \\
t(0)=t_s(0),
\end{cases}
\end{equation}

where $\varphi(r,t):=e^{\lambda -\nu}$. For a
problem of this kind, in which the initial point is singular (the
function $\varphi$ is not defined at $(0,t_s(0))$) no general
results of existence/non existence are known.

In what follows we ere going to consider {\it sub} and {\it
super} solutions of \eqref{f}. We recall that a function $y_0(r)$
is called a subsolution (respectively supersolution) of an
ordinary differential equation of the kind $y'=f(r,y)$ if it
satisfies $y_0'\leq f(r,y_0)$ (respectively $\geq$). In
\cite{GGM} we have shown the following results:

\begin{lem}\label{lem:let}
If the weak energy condition holds and the radial stress is
non-negative then the apparent horizon $t_h(r)$ is a subsolution
of \eqref{f}.
\end{lem}

\begin{teo}\label{teo:let}
If the weak energy condition holds and $\partial\varphi /
\partial t <0$ in a neighborhood of $(0,t_s(0))$ the singularity
is covered.
\end{teo}

\begin{rem}\label{rem:let}
The above results hold also in presence of radial tensions (i.e.
negative radial stresses) provided that $p_r$ satisfies, near the
centre, the bound $8\pi R^2 p_r\ge -1.$
\end{rem}

We take now the opposite point of view: we search for conditions
for naked singularity formation.

\begin{defin}\label{def:super}
A curve $t_+(r)$ is called a {\it sub-horizon supersolution (SHS)
} if
\begin{equation}\label{eq:super}
t_+(0)=t_h(0),\quad t_+(r)<t_h(r) \ \quad t_+'(r)\geq \phi (r,t_+
(r))\ \forall r>0 .
\end{equation}
\end{defin}

\begin{teo}\label{teo:nak}
The singularity is naked if and only if a SHS exists..
\end{teo}

\begin{proof}
If the singularity is naked the singular geodesic (the solution)
is also a supersolution. We now proof sufficiency. Take a point
$(r_0,t_0)$ in the region $S =\{(r,t)\,:\, r>0,
t_+(r)<t<t_h(r)\}$. At this point the (regular) Cauchy problem for
$\varphi$ admits a unique local solution $t_g(r)$. Now the
extension of this solution in the past cannot escape from $S$
since either it would cross the supersolution from above (leading
to a regular geodesic, eventually escaping from the central point
at a time prior to singularity formation) or it would cross the
subsolution from below. Thus it must extend back to the
singularity with $\lim_{r\to 0^+}t_g(r) =t_s(0)$.
\end{proof}

\begin{rem}\label{rem:conv}
It is easy to show that, if a SHS exists, then
$\frac{\partial\varphi}{\partial t}$ cannot be strictly negative
in a neighborhood of $(0,t_0)$, in accordance with theorem
\ref{teo:let}.
\end{rem}

The above analysis is limited to {\it radial} null geo\-de\-sics.
However, it can be shown that, if no radial null geodesic escapes
from the singularity, then no null geodesic escapes at all. In
other words, we show that a radially censored singularity is
censored \cite{cmp}. In the special case of dust spacetimes, this
result has been was first given by Nolan and Mena \cite{Nolan}.

\begin{teo}\label{teo:nonrad}
A radially censored central singularity is censored.
\end{teo}

\begin{proof}
Suppose, by contradiction, that the center is radially censored
but that there exist a singular non radial null geodesics
(${\tilde t}(r)$, say). This curve satisfies to
\begin{equation}\label{eq:nonrad}
\frac{\text d{\tilde t}}{\text
dr}=\sqrt{e^{2\lambda-2\nu}+e^{-2\nu}L^2/R^2}\geq \varphi(r,\tilde t) ,
\end{equation}
where $L^2$  is the conserved angular momentum. Thus ${\tilde
t}(r)$ is a supersolution of the null radial geodesic equation.
Obviously, ${\tilde t}(r) < t_h(r)$ so that ${\tilde t}(r)$ is a
SHS. As a consequence of Theorem \ref{teo:nak}, there exist a
singular null {\it radial} geodesic i.e. a contradiction.
\end{proof}

\end{section}

\begin{section}{Old and new examples}\label{sec:ex}

Comoving coordinates $r,t$ are extremely useful in dealing with
gravitational collapse because of the transparent physical meaning
of the comoving time. We are, however, going to use in the present
section another system of coordinates, the so-called area-radius
coordinates, which were first introduced by Ori \cite{Ori} to
study charged dust, and then successfully applied to other models
of gravitational collapse (see e.g. \cite{ha,m2}). These
coordinates prove extremely useful for technical purposes.

In area-radius coordinates the comoving time is replaced by $R$.
The velocity field of the material $v^\mu=e^{-\nu}\delta^\mu_t$
transforms to $v^\lambda=e^{-\nu}\dot R \delta^\lambda_R$ and
therefore the transformed metric, although non diagonal, is still
comoving. In a recent paper, we have found the following class of
solutions of the Einstein field equations in area-radius
coordinates\cite{cmp}:

\begin{equation}\label{eq:dsori}
\text ds^2=-\left(1-\frac{2\Psi}R\right)G^2\text dr^2 + 2 G\,
\frac Yu \text dR\text dr -\frac 1{u^2} \text dR^2 +R^2 (\text
d\theta^2 + \sin^2\theta\,\text d\varphi^2)\ .
\end{equation}

In the above formulae, the two functions $\Psi(r,R)$ and $Y(r,R)$
are arbitrary (positive) functions while
\begin{equation}
u^2=Y^2+\frac{2\Psi}R -1.
\end{equation}
and the function $G$ is given in terms of a quadrature:
\begin{equation}
G(r,R)=\int_R^r\frac{1}{Y(r,\sigma)}\diff{(1/u)}r(r,\sigma)\,\text
d\sigma+\frac{1}{Y(r,r)u(r,r)}.
\end{equation}
The matter distribution of the source is described in terms of
energy density, radial and tangential stresses as follows (a comma
denotes partial derivative):
\begin{align}
&\epsilon=\frac{\Psi_{,r}}{4\pi R^2 Y uG}+ \frac{\Psi_{,R}}{4\pi
R^2}
,\label{eq:eps}\\
&p_r=-\frac{\Psi_{,R}}{4\pi R^2},\qquad p_t=-\frac{1}{8\pi
RuG}\left(\frac{\Psi_{,rR}}{Y}-\frac{\Psi_{,r}\,Y_{,R}}{Y^2}
\right) -\frac{\Psi_{,RR}}{8\pi R}.\label{eq:press}
\end{align}
Special, very well known sub-cases of the metrics above are:\par
1) The dust (Tolman-Bondi) spacetimes. All stresses vanish, energy
density equals the matter density, and this implies $\Psi_{,R}=0$
and $Y=E\Psi_{,r}$.
\par
2) The general exact solution with vanishing radial stresses.
Vanishing of $p_r$ implies $\Psi_{,R}=0$, while $Y$ depends, in
general, also on $R$. The properties of these solutions have been
widely discussed in \cite{m1}.
\par

We are now going to show that Theorem \ref{teo:nak} allows for a
complete classification of all the naked singularities contained
in the metrics \eqref{eq:dsori}. Since in turn such metrics
contain all\footnote{As far as we know the unique exceptions are
the Vaidia spacetimes \cite{jbook}, which however can be easily
accommodated in the present formalism, and the case of
self-similar perfect fluids, where however the field equations
become ordinary differential equations and the standard techniques
of dynamical systems can be applied \cite{car}.} the special cases
for which censorship has already been investigated, we also get a
unified treatment of all these  - quite scattered around -
results. In every such cases, the nature of the endstates has been
obtained after a considerable amount of joint work by several
authors, sometimes developing original mathematical techniques
such as the so called root equation technique formalized by Joshi
and Dwidedi \cite{JoshiCMP}.

The arbitrary functions $\Psi$ and $Y$ must be chosen in such a
way that the metric admits a regular center prior to singularity
formation and that the weak energy condition is satisfied. It is
not difficult to check that this implies constraints on the first
non-vanishing terms of the two arbitrary functions near the
center. It is, however, useful for technical reasons to express
such constraints in an equivalent way using $Ru^2$ and $Y^2$, as
follows:
\begin{equation}\label{eq:H}
Ru^2=\sum_{i+j=3} h_{ij} r^i R^j+\sum_{i+j=3+p} h_{ij} r^i
R^j+\ldots,
\end{equation}
\begin{equation}\label{eq:Y}
Y^2(r,R)=1+f_lr^l+\sum_{\substack{i+j=q+1 \\ j>0}} k_{ij} r^i
R^j+\ldots.
\end{equation}
where $h_{30}$ (that we will call $\alpha$ hereafter) is a
strictly positive quantity, $l\geq 2$ and dots denote higher
orders terms.

It can now be shown that the following Taylor expansion of
$G(r,0)$ near the centre follows:
\begin{equation}\label{eq:G0exp}
G(r,0)=p\,a\,r^{p-1}-b r^q,
\end{equation}
where, defined
\begin{subequations}
\begin{align}
&P(\tau)=\sum_{i+j=3}h_{ij}\tau^j,
&Q(\tau)=-\sum_{i+j=3+p}h_{ij}\tau^j,\label{eq:PQ}\\
&S(\tau)=\sum_{\substack{i+j=q+1 \\ j>0}}
k_{ij}(1-\tau^j),
&T(\tau)=\sum_{i+j=3}i\,h_{ij}\tau^j,\label{eq:ST}
\end{align}
\end{subequations}
it is
\begin{equation}\label{eq:ab}
a=\int_0^1\frac{Q(\tau)\sqrt\tau}{2 P(\tau)^{3/2}}\,\text
d\tau,\qquad
b=\int_0^1 S(\tau)
\frac{\sqrt\tau\,T(\tau)}
{2P(\tau)^{3/2}} \,\text d\tau.
\end{equation}
Thus, we finally have
\begin{equation}\label{eq:G0} G(r,0)=\xi
r^{n-1}+\ldots,
\end{equation}
where $\xi>0$ and $n$ is a positive integer, namely the smallest
between $p$ and $q+1$, although very special cases -- where the
two terms balance each other and one must compute higher order
terms -- can conceived. In particular, the second term is related
to the acceleration of the matter flow lines and vanishes if this
does.

The folllowing can now be proved \cite{cmp}:

\begin{teo}\label{teo:main}
In the spacetime described by the metric \eqref{eq:dsori}, a SHS
exists - and therefore the singularity forming at $R=r=0$ is naked
- if and only if $n=1$, $n=2$, or $n=3$ and $\xi>\alpha
\xi_{\text{c}}$ where $\xi_c=\frac{26+15\sqrt{3}}{2}$.
\end{teo}

In what follows, we briefly discuss the cases already known as
well as some new ones.

\subsection{Dust clouds}\label{ex:dust}
In dust models both $\Psi$ and $Y=$ depend only on $r$. The
acceleration vanishes and therefore the index $n=p$. The solution
can be given in explicit form. The spectrum of endstates for these
models is very well known and was first calculated in full
generality by Singh and Joshi \cite{sj}. However, our unifying
framework somewhat changes the way in which the physical content
of this structure must be understood, so that we discuss it
briefly.

Introduce the Taylor expansions
\begin{equation}\label{serve12}
\Psi=F_0 r^3 + F_{\tilde{n}} r^{3+\tilde{n}}+\ldots,\quad
Y^2(r)=1+f_2 r^2+f_{m+2} r^{2+m}+\ldots.
\end{equation}
Considering first the so-called marginally bound case (this case
has been recently re-analyzed {\it via} o.d.e. techniques in
\cite{GM,Nolan}) $Y\equiv 1$, we have that the index
$n=p=\tilde{n}$ and $\alpha=2F_0,\, P(\tau)=2 F_0,\,
Q(\tau)=-2F_n$. The central singularity is naked if $n=1,2$,
censored for $n>4$. In the critical case $n=3$ the transition is
governed by the parameter $a$ \eqref{eq:ab}:
\[
a=-\int_0^1\frac{\sqrt\tau\,\, 2F_3}{2 (2 F_0)^{3/2}}\,\text
d\tau=-\frac{2F_3}{3(2 F_0)^{3/2}}>\frac{2F_0}3\xi_c.
\]
In the general (non marginally bound) case, the function $Y$ is no
longer constant. As a consequence, the expansion of the initial
density and velocity mix up to produce the index $n$ and the
transitional behavior. The singularity is naked if
$(F_1,f_3)\not=0$, or if $(F_1,f_3)=0$ but $(F_2,f_4)\not=0$,
corresponding respectively to the cases $n=1,2$. The critical
case occurs when $(F_1,f_3)=(F_2,f_4)=0$, but $(F_3,f_5)\not=0$.
Since $ \alpha=2F_0,\,P(\tau)=2F_0+f_2\tau,\,
Q(\tau)=-(2F_3+f_5\tau)$, for the singularity to be
naked\footnote{In \cite{sj} a quantity called $Q_3$ plays the
role of $a$, and its expression should be corrected - formula
(36) in \cite{sj} - to $Q_3=3a\sqrt{F_0}$, with $a$ given above.}
it must be
\begin{multline*}
a=-\int_0^1\frac{(2F_3+f_5\tau)\sqrt\tau}{2(2F_0+f_2\tau)^{3/2}}\,\text
d\tau=\\
=\frac 1{\sqrt{2F_0}}\left[\Gamma
(-\frac{f_2}{2F_0})\left(\frac{F_3}{F_0}-\frac
32\,\frac{f_5}{f_2}\right)-\frac 1{\sqrt{1+\frac{f_2}{2F_0}}}
\left(\frac{F_3}{F_0}-\frac{f_5}{f_2}\right)\right]
>\frac {2F_0}{3}\xi_c,
\end{multline*}
where the function $\Gamma$ is defined as
\begin{equation}\label{eq:Gy}
\Gamma(y)=
\begin{cases}
-\frac{\arcsin\!\text
h\sqrt{-y}}{(-y)^{3/2}}-\frac{\sqrt{1-y}}y,&\text{for\ } y<0,\\
\frac 23, &\text{for\ } y=0,\\
\frac{\arcsin\sqrt{y}}{y^{3/2}}-\frac{\sqrt{1-y}}y,&\text{for\
}y>0.
\end{cases}
\end{equation}
Finally, if $(F_1,f_3)=(F_2,f_4)=(F_3,f_5)=0$ the singularity is
censored.

To discuss the physical meaning of this structure we start from an
interesting example given by Dwidedi and Joshi \cite{JD}.

Consider a mass distribution which generates a initially
homogeneous energy density ($\Psi(r)=F_0 r^3$). Now, keeping fixed
$\Psi$, consider different initial velocities. If the cloud starts
collapsing from rest the resulting spacetime is nothing but the
"paradigm" of blackhole formation, namely, the Oppenheimer-Snyder
solution. However, it suffices a fifth-order $Y^2$, say
$Y^2(r)=1+f_0 r^2+f_5 r^5$, to change drastically the behavior of
the system. In fact, the fifth order term generates the critical
case.

What actually happens is, that the mechanism controlling existence
of SHS's depends only on the value of $n$, {\it independently}
from the details of its physical origin. The mechanism works as a
"kibbling machine" looking at just one term of the Taylor
expansion of just function near the center. It accepts any
(physically sound) input and returns an output which depends on
the value of an integer: we can call it a {\it n-machine }.

As we shall see in next example, also the equation of state plays
a similar role of "just an input" for the n-machine.

\subsection{Vanishing radial stresses}\label{subsec:vrs}

These solutions are characterized by a mass depending only on $r$,
but by a function $Y^2$ which contains the contribution of the
internal elastic energy and thus depends also on $R$. As a
consequence, the metric cannot be given in explicit form in
general (a notable exception exists however \cite{ha}).

For simplicity, we consider here only the marginally bound case.
To describe the data it is convenient to consider the Taylor
expansion of the mass and of the function $g\equiv Y^2-1$. The
constraints of physical acceptability then imply \cite{m1}:
\[
\Psi(r)=F_0 r^3+F_{\tilde n} r^{3+\tilde n}+...\ , \
g(r,R)=-\frac{\beta_k r^k(r-R)^2}{1+\beta_k r^k (r-R)^2}+...,
\]
where $k\geq 1$. Although the solutions are accelerating, it can
be easily shown that acceleration terms do not enter into the
first non-vanishing term of formula \eqref{eq:G0}. We thus need to
consider the Taylor expansion of $Ru^2$, given to lowest orders by
$2 F_0 r^3 + 2 F_{\tilde n} r^{3+\tilde n}-\beta_k r^k\,R\,
(r^2+R^2-2rR)$. Since in general the solution is known only by
quadratures, when the spectrum of endstates was first calculated
in full generality \cite{gjm} the Taylor expansion was used under
the integral using approximation techniques near the center
(delicate estimates coming from Lebesgue dominated convergence
theorem are required). The endstates {\it look like} to depend on
two parameters $\tilde n,k$, the second of them coming exclusively
from the non-trivial equation of state. However, again, if we use
the n-machine of Theorem \ref{teo:main} we see that the mechanism
has no interest at all in the physical source of the parameter
$n$: a naked singularity ($n=1,2$) can originate from $k=1,2$ {\it
or} from $\tilde n =1,2$. The critical case always occurs at $n=3$
and what changes is only the value of the critical parameter,
since it must be $a_{k\tilde n}>\frac{2F_0}{3}\xi_c$,
where\footnote{In \cite{gjm} a multiplying factor is missing.}
\begin{multline*}
a_{k \tilde n}=\int_0^1\frac{-2F_3\delta_{\tilde n 3
}\tau^{1/2}+\beta_3 \delta_{k3}
(\tau^{3/2}-2\tau^{5/2}+\tau^{7/2})}{2(2F_0)^{3/2}}\,\text
d\tau=\\=\frac 1{(2F_0)^{3/2}}\left[-\frac 23 F_3\delta_{\tilde n
3 }+\frac 8{315}\beta_3\delta_{k 3 }\right],
\end{multline*}

\subsection{Acceleration free non-dust solutions}

In recent years, spacetimes with cosmological ("lambda") term have
attracted a renewed interest both from the astrophysical point of
view, since recent observations of high-redshift type Ia
supernovae suggest a non-vanishing value of lambda, and from the
theoretical point of view, after the proposal of the so-called
Ads-Cft correspondence in string theory. The unique model of
gravitational collapse with lambda term available so far is the
Tolman--Bondi--de Sitter (TBdS) spacetime, describing the collapse
of spherical dust (within the framework of the present paper,
these solutions are obtained choosing $\Psi(r,R)=F(r)+\frac\Lambda
6 R^3$). The spectrum of endstates for these solutions has
recently been obtained \cite{Gon}.

One way to construct new collapsing solutions with lambda term is
to choose a function of the type $\Psi(r,R)=F(r)+M(R)\,R^3$, with
$M(R)$ not necessarily constant as in TBdS. It is easy to check
that, in order to satisfy the requirements of physical
reasonableness, one must have $\Psi(r,R)=F_0 r^3 +
F_{\tilde n} r^{3+\tilde n} + M_0R^3+M_k R^{k+3}+...$. Then, if
one among $(F_1,F_2,M_1,M_2)$ is non zero the singularity is
naked. If otherwise they all vanish, and one among $(F_3,M_3)$
does not, we must compute the parameter $a$ \eqref{eq:ab}:
\[
a=-\int_0^1\frac{\sqrt\tau(F_3+M_3\tau^3)}{(2F_0+2M_0
\tau^3)^{3/2}}\,\text d\tau=
\frac{2}{3(2F_0)^{3/2}}\left[M_3\,\Gamma(-\frac{M_0}{F_0})-
\frac{F_3+M_3}{\sqrt{1+\frac{M_0}{F_0}}} \right],
\]
where $\Gamma$ is given in \eqref{eq:Gy}. The singularity is naked
if $a>\frac{2 F_0}3 \xi_c$. In all other cases the singularity
is censored.\footnote{In the particular case when all the $M_k$
vanish for $k>0$, we recover Tolman--Bondi--de Sitter spacetime. The
condition for nakedness at the transition reduces to
$\frac{-F_3}{(2F_0)^{5/2}\sqrt{\frac{M_0}{F_0}+1}}>\frac{\xi_c}2$,
slightly correcting a wrong value given in \cite{Gon}.} Once
again, the physical parameters of the collapse (in this case,
there is also the cosmological constant) are mixed up by the
n-machine.

\end{section}

\begin{section}{Discussion}\label{sec:d}

Starting from the piooneristic work by Eardley and Smarr \cite{es}
and Christodoulou \cite{C1} on dust clouds, the number of papers
containing analysis of examples of naked singularities in
spherically symmetric spacetimes can, up to now, be estimated to
more than one hundred. Authors have tried to make clear the
connection between, on one side, the final states of the collapse
and, on the other side, the choice of the initial data and of the
matter models. To do this, a mathematical framework was developed
(the so called root equation technique) which allows investigation
of the behavior of solutions of a differential equation at a
singular point. To be applicable, this framework requires explicit
knowledge of the solution of the Einstein field equations in which
the geodesic motion is studied: it cannot be applied if only a
part of the field equations are integrated, and can be applied
only with enormous technical efforts if the solution is known only
by quadratures. It is anyway obvious that, due to the non-linear
character of the Einstein equations one cannot have any choice of
giving general answer to the censorship problem, even in spherical
symmetry, using {\it only} exact solutions.

So motivated, we started re-considering the censorship problem in
spherical symmetry in an effort of going beyond the need for exact
solutions. Essentially, the idea was to {\it rely} as much as
possible on the "differential level" - the fact that the metric
satisfies to the field equations - without resorting to the
fortunate case of explicit solutions. This approach leads
naturally to the application of comparison techniques in o.d.e.
since the physical properties of the apparent horizon have as
mathematical counterpart the fact that this horizon is a
sub-solution of the geodesic motion \cite{GGM}. Here we presented
a new framework for the investigation of censorship, based on such
ideas. The framework can be easily applied when the exact solution
is known only up to quadratures, and this allowed us to give a
unified view of (virtually) all the examples of naked
singularities which are scattered around in the literature, as
well as to produce new examples. The application of such
techniques allows, however, also to investigate cases in which
the metric is not known at all, such as, for instance, the case
of barotropic perfect fluids.

\end{section}

\end{document}